\documentclass[journal=jacsat,manuscript=article,email=false]{achemso}
\usepackage[version=3]{mhchem} 
\usepackage{xcolor}
\usepackage[normalem]{ulem} 

\usepackage{multirow}
\usepackage{tablefootnote}
\usepackage{caption}
\usepackage{subcaption}
\usepackage{multicol}
\usepackage{array}
\usepackage{amsmath}
\usepackage{amssymb}

\usepackage{upgreek}

\usepackage{graphicx} 
\usepackage{float} 
\usepackage{listings} 
\usepackage{xcolor}

\usepackage{tikz}
\title{Transfer Learning for Multi-material Classification of Transition Metal Dichalcogenides with Atomic Force Microscopy}
\author{Isaiah A. Moses}
\affiliation[MRI]{Materials Research Institute, The Pennsylvania State University, University Park, PA 16802}
\author{Wesley F. Reinhart*}
\affiliation[MTSE]{Department of Materials Science and Engineering, The Pennsylvania State University, University Park, PA 16802}
\alsoaffiliation[ICDS]{Institute for Computational and Data Sciences, The Pennsylvania State University,
University Park, PA 16802}

\begin{document}

\maketitle
\let\thefootnote\relax\footnote{*Corresponding Author (reinhart@psu.edu)}

\begin{abstract}
Deep learning models based on atomic force microscopy (AFM) enhance efficiency in inverse design and characterization of materials. However, the limited and imbalanced data of experimental materials that are typically available is a major challenge. Also important is the need to interpret trained models, which are normally complex enough to be uninterpretable by humans. Here, we present a systemic evaluation of transfer learning strategies to accommodate low-data scenarios in materials synthesis and a model latent feature analysis to draw connections to the human-interpretable characteristics of the samples. While we imagine this framework can be used in downstream analysis tasks such as quantitative characterization, we demonstrate the strategies on a multi-material classification task for which the ground truth labels are readily available. Our models show accurate predictions in five classes of transition metal dichalcogenides (TMDs) (\ce{MoS2}, \ce{WS2}, \ce{WSe2},\ce{MoSe2}, and \ce{Mo-WSe2}) with up to 89\% accuracy on held-out test samples. Analysis of the latent features reveals a correlation with physical characteristics such as grain density, Difference of Gaussian blob, and local variation. The transfer learning optimization modality and the exploration of the correlation between the latent and physical features provide important frameworks that can be applied to other classes of materials beyond TMDs to enhance the models' performance and explainability which can accelerate the inverse design of materials for technological applications.
\end{abstract}

\paragraph{Keywords:} Transition metal dichalcogenides, Atomic force microscopy, Computer vision, Model interpretation, Transfer learning.

\section{Introduction}
Transition metal dichalcogenides (TMDs) are an important class of two-dimensional materials with interesting properties that promise great potential for applications in technological applications.\cite{jayachandran2024three,
 upadhyay2021recent, joseph2023review, kirubasankar2022atomic, lau2023dielectrics} The materials are being explored for applications in areas including, energy,\cite{pawar2018heterojunction, beckmann2024self, park20192d} optoelectronics,\cite{jayachandran2024three, 27schranghamer2023ultrascaled, gong2016thin, sebastian2021benchmarking, kwon2024200, liu2021wafer, hong2021wafer} and sensing.\cite{ping2017recent, beckmann2024self, li2022recent, zappa2017molybdenum} 
Different characterization techniques are used to determine the quality and properties of the grown TMDs. Some of these focus on the determination of the chemical signature and characteristics (such as X-ray photoelectron spectroscopy (XPS), mass spectrometry, and Auger Electron Spectroscopy (AES)),\cite{nolot2020line, jernigan2020electronic, chubarov2018plane, li2020surfactant} while others are related to the structural properties of the materials (such as X-ray diffraction (XRD), Raman spectroscopy, scanning and transmission electron microscopy,\cite{1eichfeld2015highly, kang2015high} and Atomic Force Microscopy (AFM)),\cite{lu2016atomic, huang2014large, lin2014direct, 5zhang2016influence} and still others on the physical properties (such as electrical measurements, mechanical testing, and optical spectroscopy).\cite{wang2016interlayer, wang2017probing, zeng2015optical} Irrespective of the characterization method, the materials are a function of their chemical composition and the growth and after-growth conditions.\cite{wang2018atomic} AFM, a scanning probe microscopy technique, provides topographic and various surface properties of materials at the nanoscale.\cite{rugar1990atomic, giessibl2003advances, zhang2018atomic} The AFM images, containing a high-dimension representation of the materials, are analyzed to determine the quality and properties of the grown materials. 

In data-driven design and exploration of materials, deep learning models (DL), including computer vision, have been widely used to develop models for the analysis and classification of different classes of materials based on AFM.\cite{moses2024quantitative, moses2024crystal, stuckner2022microstructure, 10borodinov2020machine, p13oinonen2022molecule, 12alldritt2020automated, gordon2020automated, carracedo2021deep, ziatdinov2022atomai}
DL and machine learning (ML) minimize the amount of expert time required for materials characterization, analysis, detection, and classifications. Their deployment eradicates biases inherent in manual postprocessing of materials characterization data. They also provide tools that enhance efficiency and an avenue for high throughput growth and characterization of materials with predefined properties for technological applications.\cite{21tang2020machine, 22beckham2022machine, lu2022machine, frey2019prediction, ryu2022understanding}

A major challenge in deploying DL in materials science is the large amount of data typically required in training the models.\cite{moses2024quantitative, carracedo2021deep, moses2022accelerating, 12alldritt2020automated} While simulated data are being used in some studies,\cite{carracedo2021deep, lee2022stem, 12alldritt2020automated} the experimentally grown materials data are limited and pose a significant bottleneck to the data-driven design and exploration of materials, as the former at best is only an approximation to the latter. Besides the challenge of insufficient data, we also have the associated limitations of imbalanced training data. Materials with desirable properties are more likely to be grown in larger quantities than those without (Table~\ref{tab:data}). Also, it is common to have a material grown with parameters that result in the quality and property of interest, while less favorable parameters are rarely used.\cite{moses2024quantitative} 

Transfer learning is the use of knowledge gained in training a source, usually larger, data set in target data to improve the generalization of the model trained on the latter.\cite{pan2009survey} The weights of the model trained on the source data, pre-trained weights, are used, either by fine-tuning them on the target data or by using them to extract the features of the target data, and then training shallow or unsupervised models. It has been a key tool in addressing the challenge of limited data in DL for data-driven materials characterization.\cite{stuckner2022microstructure, moses2024crystal, moses2024quantitative} For instance, VGG16 pre-trained on ImageNet has been used to extract features to classify microstructure images\cite{kitahara2018microstructure} and TEM images of carbon nanomaterials.\cite{luo2021transfer} Similarly, semantic segmentation has been performed with diverse data sets using several computer vision architectures pre-trained on ImageNet and MicroNet.\cite{stuckner2022microstructure} More recently, we used the ResNet18 model pre-trained on ImageNet and another on MicroNet to perform feature extraction and fine-tuning methods for the classification of \ce{MoS2} AFM images based on their growth temperature.\cite{moses2024quantitative} Also, we trained segmentation and regression models based on ImageNet and MicroNet pre-trained weights to predict the mono-layer crystal coverage and crystal-substrate pixel-wise classification of \ce{WSe2} AFM samples.\cite{moses2024crystal} Instead of random initialization of millions of weights that need to be trained on limited target data, models pre-trained on larger data that share some generic features with the target data can be deployed in transfer learning. A divergence between the source and target data, where common generic features are limited, could result in negative transfer learning.\cite{pan2009survey, wang2019characterizing} The optimization of the transfer learning for best performance using the TMD-AFM data as a demonstration will significantly guide their use in materials science.

Beyond taking measures that ensure trained models perform optimally and can generalize to new data, the models' interpretation is an important consideration associated with DL in materials science. To use DL models to accelerate materials discovery through high-throughput materials design and characterization, it is important to derive as much utility from the trained models as possible. It is, therefore, important to go beyond obtaining the predictions provided by the models to explore the basis and the hows of the model's decisions. For instance, images go through different layers of representation in convolutional neural network (CNN) models. Some features of a few dimensions could be obtained from the later layers, usually with fully connected nodes. A major utility is derivable if such features, known as the latent features, could be interrelated with the physical feature of the image. The growth parameters, for instance, could be related to the latent features and hence to the physical features, which provide information for an inverse design of materials.

A successful classification of the TMDs based on their AFM images will require identifying latent features peculiar to the different classes, even if they are grown under the same conditions. In addition to the class of the TMDs,  image features related to the chemical constituent are of interest as the high throughput growth of the materials for on-demand technological application is sought. While models are trained to classify materials, a crucial utility that arises is the enablement to represent the image in a few latent dimensions. Latent space from the machine learning models enables a representation of the image in a very reduced dimension compared to the original multidimensional image. An important application derivable from such representation is an easy explanation of the AFM image in terms of the specific material properties, chemical composition, and the material's growth conditions. Additionally,  an interrelationship of the latent and the physical features and the performance metrics of the materials could be analyzed.

We present deep learning models trained on AFM images (obtained by contact mode scanning) of five classes of metal-organic chemical deposition (MOCVD) grown TMDs. The materials were grown at the Pennsylvania State University 2D Crystal Consortium (2DCC)\cite{1eichfeld2015highly, 5zhang2016influence, 28trainor2022epitaxial, 15zhang2018diffusion, 4eichfeld2016controlling, huet2023mocvd, zhu2023step, nayir2023modulation, chubarov2018plane, reifsnyder2021illuminating, hickey2020formation} and hosted in the Lifetime Sample Tracking (LiST) database.\cite{Moses_Reinhart} The TMDs included in our data are \ce{MoS2}, \ce{WS2}, \ce{WSe2}, \ce{MoSe2}, and \ce{Mo-WSe2}. Two classification approaches based on target representation, multiclass (MtC) and multilabel (MtL), and different modes of transfer learning, were investigated to propose what is suitable for such analysis. We also proposed a modality for maximizing transfer learning utility in cases of imbalanced data and when models are trained on a few classes, and an update is required to account for more classes than were previously included in the original models. A significant contribution in this study also includes the determination of the discriminative latent features used by the ML models in classifying materials and their correlation with physical features. Also, the processed 1026 AFM image data (termed TMD-AFM)\cite{moses_2024_13961220} is publicly available to the scientific community.

\section{Methods}

\subsection{Data Preparation}
\begin{table}[ht]
	\centering
	\caption{The TMD-AFM data sets used in the models.}
	\label{tab:data}
		\begin{tabular}{l| c   c    c     |c} \hline
    TMD   &   train   &     val   &   test    & Total    \\ \hline
    \ce{MoS2}   &   344        &  85      &   47  &   476     \\
   \ce{WS2}   &   131         &   33      &   18  &   182     \\
   \ce{WSe2}  &   99     &   25      &       13   &  137     \\
   \ce{MoSe2} &   93     &   24      &       12      & 129    \\
   \ce{Mo-WSe2}&   70 &  18  &   14  &   102     \\ \hline
   Total    &  737     &   185      &       104      &   1026 \\ \hline
    \end{tabular}
\end{table}
\begin{figure}
     \centering
     \includegraphics[width=\textwidth]{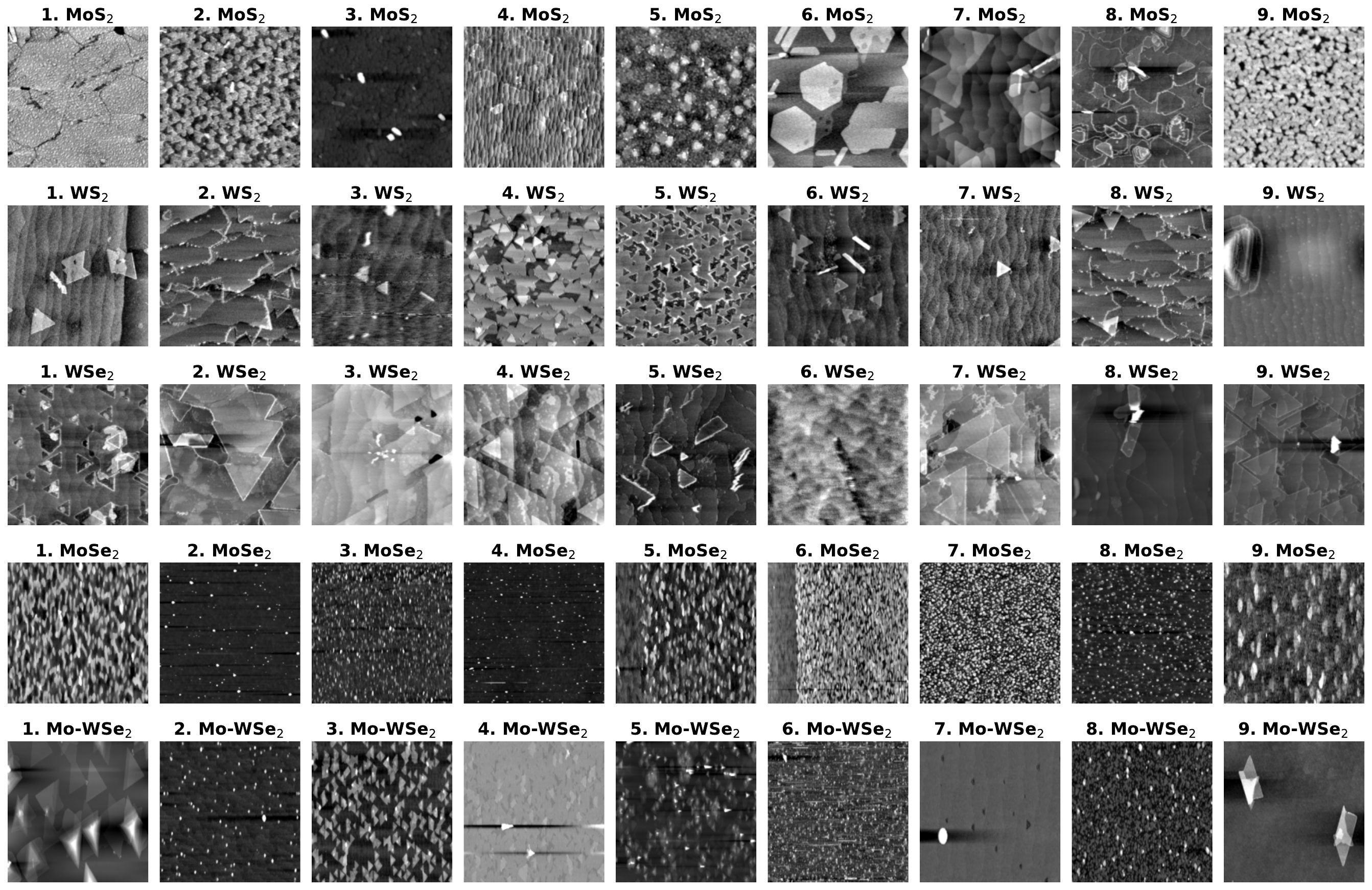}
         \caption{Samples of atomic force microscopy (AFM) images from the MOCVD grown TMDs used in the study.}
         \label{fig:data}
\end{figure}

We have used AFM images of TMDs grown using MOCVD at the 2DCC of Pennsylvania State University. The data is stored in the LiST database of the 2DCC. The data we have used consists of five different TMDs, \ce{MoS2}, \ce{WS2}, \ce{WSe2}, \ce{MoSe2}, and \ce{Mo-WSe2}. The raw SPM files of the data were retrieved and processed to obtain the AFM images. The TMDs were grown with different parameters, including temperature, pressure, time, precursor flux, etc. Additionally, diverse imaging conditions were used to obtain the data. For instance, images were taken from the center and edges of the wafer and at varied resolutions. This results in multiple images per sample of the TMD grown. For consistency in the imaging conditions, we have mostly used images taken from the center of the wafer. Exceptions are the \ce{MoSe2} and \ce{Mo-WSe2} which have fewer samples, hence their images taken from the edges of the wafer were included. For all the TMDs, only $1 \mathrm{\upmu m} \times 1 \mathrm{\upmu m}$ images taken at the center of the $5 \mathrm{\upmu m} \times 5 \mathrm{\upmu m}$ AFM scans were included in our data.

The total data, consisting of 1026 AFM images of TMDs, were split into train, validation, and test sets (Table \ref{tab:data}). Sample images of the consistent TMDs are shown in Figure \ref{fig:data}. It was ensured that a given sample was assigned to only one of these splits, even though multiple images of each sample were present (i.e., splitting by sample ID). For instance, images from a given sample represented in the train set cannot be in the test set. This ensures unique samples are present in each data set to avoid data leakage. About 10\% of the data was held out for testing, while five-fold cross-validation was carried out using the remaining data.

\subsection{Supervised learning}

\begin{figure}
     \centering
     \includegraphics[width=\textwidth]{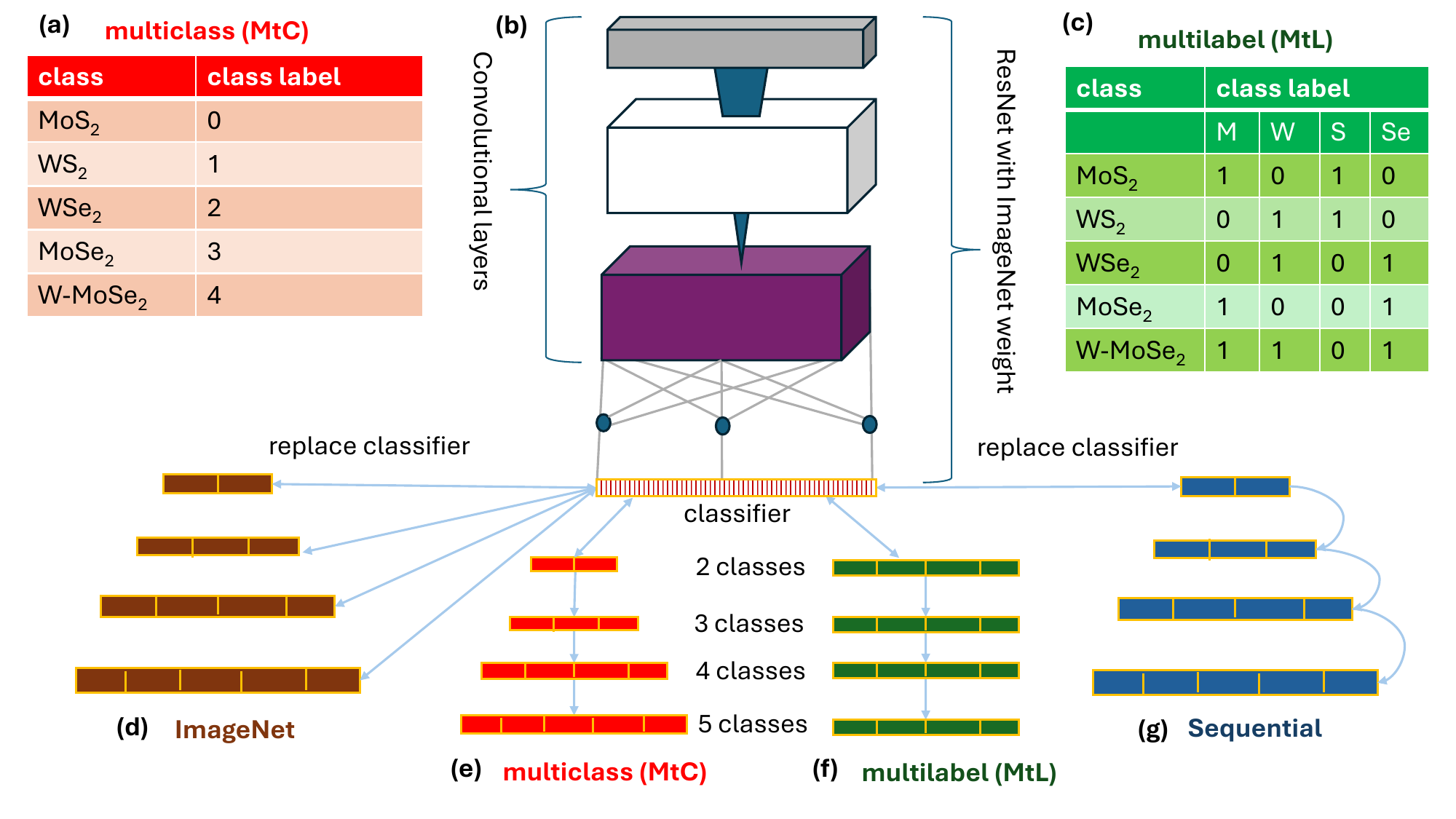}
         \caption{The target representation and transfer learning schemes. (a) and (c) are the MtC and MtL schemes, respectively. (b) is a schematic of ResNet architecture showing the convolutional and classifier layers. (d) and (g) show the two schemes of transfer learning, ImageNet, and Sequential, respectively. In ImageNet, the ImageNet weights are fine-tuned separately in classifying two, three, four, or five classes of TMDs. In Sequential, sequential training is adopted. We continue with the previous fine-tuned weights from fewer TMD classes (for instance, fine-tuning on three TMDs classification using the fine-tuned weights of two TMDs classification). Panels e and f show the number of nodes in the classifier layer for the MtC and MtL, respectively.}
         \label{fig:schemes}
\end{figure}

The award-winning ResNet architectures~\cite{ResNethe2015deep} were used in our study since they provide a residual learning framework that eases the training of very deep networks. We used the implementation in \texttt{torchvision}~\cite{torchvision}, part of the \texttt{pytorch}~\cite{PYTORCHpaszke2019pytorch} software ecosystem. The ImageNet classifier was replaced with two fully connected layers. The number of nodes for the first layer is 2048 for ResNet50, 101, and 152, while ResNet18 and 34 have 512 nodes. The second layer for all the architectures has 100 nodes, followed by an output layer. A ReLU activation function and a dropout are between the two fully connected layers. The dropout, learning rate, and batch size were optimized using grid search. 

We first explore different modes of transfer learning, the fine-tuning where the whole pre-trained model was fine-tuned on our data, and feature extraction where features were extracted from the pre-trained model and used to train a multilayer perceptron (MLP), were explored. The smallest ResNet architecture, ResNet18, pre-trained on the ImageNet data was examined for the transfer learning. Additionally, as a baseline, the ResNet18 architecture was trained from scratch with randomly initialized weights. A \texttt{pytorch}~\cite{PYTORCHpaszke2019pytorch} implementation weighted random sampler was used to mitigate the imbalance in the training data.
The data were normalized using the mean and the standard deviation of the ImageNet data, while vertical and horizontal flips at 50\% were applied to the training data on the fly.

We used two different schemes in representing our target, the multiclass (MtC) and multilabel (MtL) (Figure~\ref{fig:schemes}). In MtC, each TMD class is represented as an independent class with a scalar label, 0, 1, 2, 3, or 4. In MtL, each class is labeled using one-hot encoding with a four-dimensional vector to account for the presence or absence of the constituent atoms in the TMDs represented in our data ([M, W, S, Se] in that order). Each component indicates an atom is present (1) or absent (0). For instance, for \ce{MoS2} containing \ce{Mo} and \ce{S}, the label is [1, 0, 1, 0]. The loss functions used for the MtC and MtL classifications are CrossEntropy and Binary CrossEntropy (BCE), respectively, while the Adam optimizer was used in both cases.
The metrics used are defined in the Supporting Information.

To quantify the performance of our models while training, the F1 score, and the MSE were used as metrics to determine what model to save while training. After an epoch of training, the previously saved model (if any) is replaced with the current model if and only if the F1 score and MSE value on the validation improve compared to the previously saved model. The model was trained for 60 epochs. However, the best model saved is usually obtained before 40 epochs are reached, as no further improvement in the validation score is recorded beyond that.

To minimize bias in assessing the model performance, a five-fold cross-validation was used in the training. In five-fold cross-validation, the training set (train and val set in Table \ref{tab:data}) is split into five folds. A fold is used for validation and the remaining fourfolds for training the model. Therefore, five models (with the same hyperparameters), with a unique fold serving as a validation set while the remaining for training, were trained and evaluated.

\subsection{Model interpretation}
To explain the trained model, class activation maps (CAM) of the different TMD classes are obtained. The CAM implementation by Zhou, et al\cite{CAMzhou2016cvpr} was used. After the pre-trained ResNet model is fine-tuned on our data, the feature maps of the last convolutional layer of the fine-tuned model are summed up and then normalized by dividing by the maximum value to obtain a heatmap with the same dimensions as the layer. The heatmaps are obtained for sample images from the different classes of the TMDs. The dark red region on the CAM has the highest activation for the class and, hence, is most significant for the classification.

\subsection{Latent space analysis}
The interrelationship of the latent and physical image features was analyzed using Pearson's correlation coefficient and linear regressions implemented in the \texttt{scikit-learn} library version 1.2.2\cite{SCIKITLEARNpedregosa2011scikit}. In analyzing the latent space and examining how they are related to the image's physical features, 100 features of the images were extracted using the trained models. Principal component analysis (PCA) was then applied to the extracted features. PCA is a linear dimensionality reduction method that linearly transforms data into a new coordinate system such that the directions (principal components) represent the largest variance in the data.\cite{PCAdoi:10.1080/14786440109462720, PCAjolliffe2016principal} The first ten principal components, which account for about 99\% of the cumulative explained variance, were obtained. 

The material images were analyzed using several metrics, including microstructural features such as the grain size, grain density,  and Difference of Gaussian (DoG) blob,\cite{DoGlowe2004distinctive} and topographical and defect-related features such as the surface roughness, step edges, defect density, anisotropy, and local variation. The grains were determined using Otsu’s thresholding algorithm\cite{Otsu4310076} implemented in \texttt{scikit-image} library. The algorithm works by maximizing the variance between two classes of pixels to determine the threshold separating the classes. The surface roughness is measured by determining the root-mean-squared error between the Gaussian smoothed and the original image. To determine the step edges, the Canny edge detection\cite{Canny4767851} was used, and the edges were enhanced using binary dilation and erosion. Defect density is measured using the local maximum of the Gaussian smoothed image. Anisotropy is the standard deviation of the angle of gradients per the mean of the magnitudes of the gradient. Local variation is the mean of the normalized absolute difference between the Gaussian smoothed and actual image, using the Gaussian filter from the Scipy library.\cite{2020SciPy-NMeth} Others are the Gray-Level Co-occurrence Matrix (GLCM)\cite{GLCM4309314} features such as contrast, dissimilarity, homogeneity, energy, correlation, and angular second moment (ASM) implemented in \texttt{scikit-image}. Unless otherwise stated, the \texttt{scikit-image} version 0.23.2 was used for the image processing.\cite{scikit-image} More details on the image analysis implementation parameters are provided in the supporting information.

\section{Results and Discussion}

\subsection{Model Performance}
\begin{table}[ht]
	\centering
	\caption{The performance of models using pre-trained weights and training from scratch (Scratch). The ImageNet label indicates fine-tuning of ResNet18 pre-trained on ImageNet data. ImageNet* is an MLP model trained on features extracted from the average pool layer of ResNet18 pre-trained on ImageNet data.}
	\label{tab:tranfer}
		\begin{tabular}{l|    c    c | c   c} \hline
  Model&  \multicolumn{2}{c|}{\bf{accuracy (\%)}} & \multicolumn{2}{c}{\bf{F1 score}} \\ \hline
       &   train   &     val   &   train    & val    \\ \hline
   ImageNet &  \bf{95.4$\pm$2.1}         &   \bf{91.0$\pm$5.2}      &   \bf{0.96$\pm$0.02}  &   \bf{0.92$\pm$0.05}  \\ 
   ImageNet*&   89.3$\pm$2.4     &   74.5$\pm$3.1      &       0.90$\pm$0.02      &   0.74$\pm$0.03 \\ 
    Scratch &   75.4$\pm$11.7     &   67.7$\pm$12.9      &       0.78$\pm$0.12      &  0.69$\pm$0.13 \\ \hline
    \end{tabular}
\end{table}
An interesting aspect of transfer learning that has not been fully explored in the materials science context is continual learning when introducing novel materials. When considering material synthesis, wherein collections of elemental constituents can be combined in different combinations, the question naturally arises of whether new models should be created for each new material or whether some prior knowledge can be leveraged and applied to the new material. This leads to algorithmic decisions with both the learning itself as well as labeling schemes to use. The obvious choices are MtC, wherein a unique class label represents each material, and MtL, wherein a collection of elemental labels represents each material. These two strategies are illustrated in Figure~\ref{fig:schemes}. We test these two strategies with the idea that MtC is more discerning between materials while potentially leveraging less of the prior knowledge from other materials, whereas MtL uses a common labeling scheme across a series of similar elemental compositions but may learn or retain spurious information as elements combine to form different crystal structures.

We first observe the performance of our models on MtC using the fine-tuning and feature extraction approaches of transfer learning (Table \ref{tab:tranfer}). The optimum dropout, learning rate, and batch size values are 0.0, 7.5e-5, and 32, respectively, obtained by grid search. The fine-tuning yielded significantly better results than the feature extraction. While both approaches give better results than a random initialization of the weight, underscoring the importance of transfer learning, a major advantage of the fine-tuning over the feature extraction was associated with the on-the-fly augmentation benefited by the former.\cite{moses2024crystal} The following analysis is therefore based on fine-tuning.

\begin{table}[ht]
	\centering
	\caption{The performance of different pretrained ResNet architecture in classifying TMDs; \ce{MoS2}, \ce{WS2}, \ce{WSe2}, \ce{MoSe2}, and Mo-WSe$_2$. The best performance in each column is shown in bold, including ties and near-ties. The V1 and V2 in the architecture names are the weight versions 1 and 2, respectively. The ``ImageNet'' column indicates accuracy on the ImageNet classification task, as reported in the \texttt{torchvision} documentation.\cite{torchvision}}
	\label{tab:resnet}
		\begin{tabular}{l| c |  c    c | c   c} \hline
  Architecture& \bf{ImageNet} & \multicolumn{2}{c|}{\bf{accuracy (\%)}} & \multicolumn{2}{c}{\bf{F1 score}} \\ \hline
    &   &   train   &     val   &   train    & val    \\ \hline
   ResNet18$\_$V1 &  89.1 &   95.4$\pm$2.1         &   91.0$\pm$5.2      &   0.96$\pm$0.02  &   0.92$\pm$0.05    \\
   ResNet34$\_$V1 &  91.4 &   94.7$\pm$2.3     &   89.8$\pm$7.6      &      0.95$\pm$0.02   &  0.91$\pm$0.07     \\
   ResNet50$\_$V1 &  92.9  &   94.5$\pm$4.4     &   90.0$\pm$8.6      &      0.95$\pm$0.04   &  0.91$\pm$0.07     \\
   ResNet101$\_$V1 & 93.5     &   96.7$\pm$1.9     &   92.0$\pm$5.8      &      0.97$\pm$0.02   &  0.92$\pm$0.06     \\
   ResNet152$\_$V1 & 94.0     &   93.6$\pm$2.8     &   89.7$\pm$4.3      &      0.94$\pm$0.03   &  0.90$\pm$0.05     \\
   ResNet50$\_$V2 & 95.4  &   \bf{98.4$\pm$1.1}     &   \bf{93.4$\pm$5.3}    & \bf{0.99$\pm$0.01} & \bf{0.94$\pm$0.06}  \\ 
   ResNet101$\_$V2 & 95.8      &   96.5$\pm$1.4     &   91.8$\pm$4.4      &       0.97$\pm$0.01      &   0.92$\pm$0.04 \\ 
    ResNet152$\_$V2 &  \textbf{96.0}     &   \bf{97.5$\pm$1.5}     &   \bf{94.4$\pm$4.2}      &       \bf{0.98$\pm$0.01}      &  \bf{0.95$\pm$0.04} \\ \hline
    \end{tabular}
\end{table}

Different ResNet architectures with the different ImageNet pre-trained weights available (either V1 or V2, depending on the architectures in Table~\ref{tab:resnet}) were fine-tuned. This allows for comparing the different architectures and identifying the best one for our data.  The models were trained for 60 epochs, and the loss and F1 scores were evaluated after each epoch. The trained models were saved for every improvement in the loss and F1 score as the training proceeded. The final models obtained were those with the lowest loss and highest value of the F1 scores. Among the different architectures, the differences in model validation performances are minimal. Except for the ResNet101 which has the same performance with both the V1 and V2, the V2 pre-trained weights, which were obtained using a better training recipe\cite{torchvision} give better performance than the corresponding V1 weights. Interestingly, the difference between our results and the ImageNet validation accuracy is just about 2\% on average. The V2 of ResNet152 was observed to give the best validation accuracy and F1 score (Table \ref{tab:resnet}), in agreement with the ImageNet accuracy. This model was, therefore, selected for the subsequent analysis.

\begin{table}[ht]
	\centering
	\caption{Performance of pretrained ResNet152 in classifying TMDs; \ce{MoS2}, \ce{WS2}, \ce{WSe2}, \ce{MoSe2}, and Mo-WSe$_2$ using different representation of the targets, multiclass (MtC) and multilabel (MtL).}
	\label{tab:Mts}
		\begin{tabular}{l|    c  c  c | c   c c} \hline
  &  \multicolumn{3}{c|}{\bf{accuracy (\%)}} & \multicolumn{3}{c}{\bf{F1 score}} \\ \hline
       &   train   &     val   &   test & train    & val    & test \\ \hline
   
   MtC&   97.5$\pm$1.5     &  94.4$\pm$4.2  &  87.3$\pm$2.1    &       0.98$\pm$0.01     & 0.95$\pm$0.04 & 0.87$\pm$0.02 \\ 
   MtL    &   \bf{99.0$\pm$0.8}    &   \bf{95.1$\pm$5.1}    &  \bf{88.8$\pm$2.6}    &       \bf{0.99$\pm$0.01}      &  \bf{0.96$\pm$0.04} & \bf{0.90$\pm$0.02} \\ \hline
    \end{tabular}
\end{table}
\begin{figure}
     \centering
     \includegraphics[width=\textwidth]{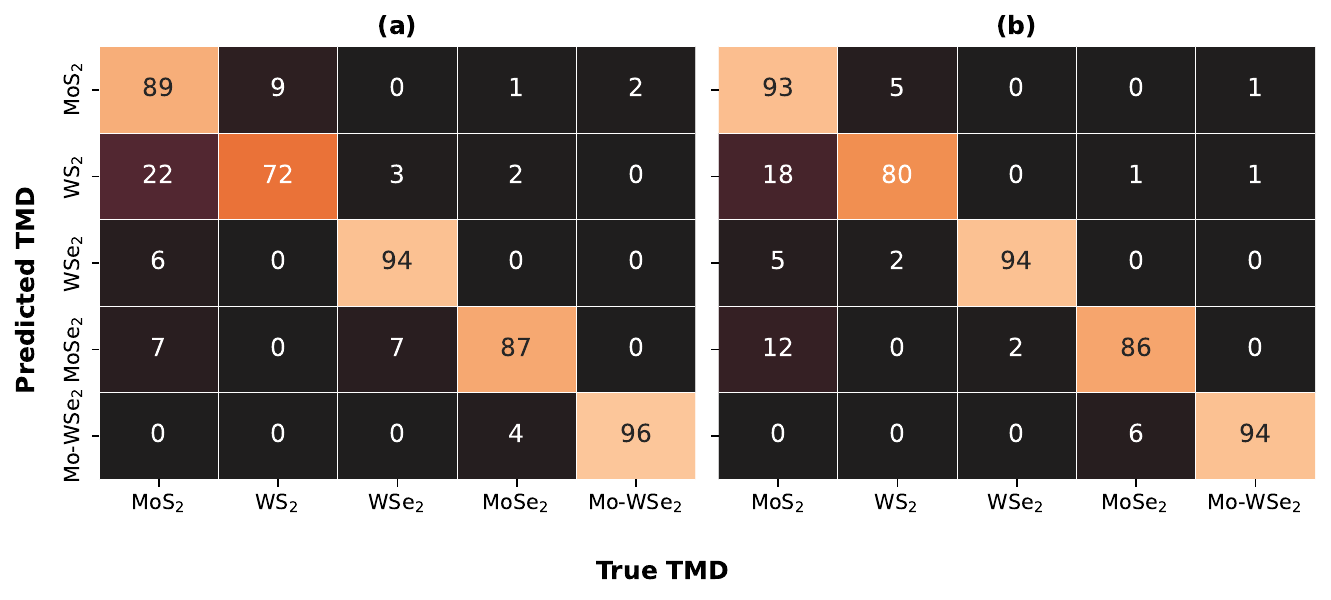}
         \caption{The confusion matrix for classifying five TMDs using (a) multiclass (MtC) and (b) multilabel (MtL) representations of the targets. The results are the percentages of the average predictions of the held-out test set using the five models from the five-fold cross-validation.}
         \label{fig:confusion}
\end{figure}

Having identified the ResNet architecture and pre-trained weights with the best performance, we compare the performance of our models with the two different training modalities: MtC label and MtL. Similar results were obtained, with the MtL having a minimal improvement over the MtC modality (Table \ref{tab:Mts}). 
While the MtC classification gives validation and test accuracies of about 94, and 87\%, respectively, the MtL gives 95 and 89\%, respectively. The high value of the F1 score established the lack of biases of our models to more represented class(s) in the data. In addition to the better performance observed, the MtL modality has the advantage of learning the constituent atom and not just the class of TMD. Therefore, to use our trained models to examine the features associated with the TMDs constituent atoms, the MtL classification training was adopted. The analyses that follow are based on MtL unless otherwise specified.

The impressive performance of the models in classifying the materials is noteworthy. This excellent performance cuts across the different data sets. In particular, the accuracy and F1 score recorded for the test set, 89\% and 0.90, respectively, which were never used in training nor in selecting the models, attests to the potential of transfer learning in general and to the specific approaches we have adopted in training the models. To gain insight into how the models perform for the different classes of TMDs, the confusion matrix is shown in Figure~\ref{fig:confusion}. The average number of classes in the held-out test set using the five models from the five-fold cross-validation are shown.

Interestingly, the few misclassifications observed are between \ce{MoS2} and \ce{WS2}. This phenomenon is observed in both the MtL and MtC classifications. To investigate the basis for the misclassification, we extracted the features of the images using pre-trained ResNet152 before fine-tuning the model on our data. A 2-dimensional representation of the data from the first two components with the highest variance shows no separation among the classes (Figure S1). However, it is observed that the \ce{MoS2} and \ce{WS2} present more overlap compared to, for instance, between the \ce{MoSe2} and \ce{WSe2}. This establishes a significant similarity in their surface morphology, at least based on filters relevant to the ImageNet classification task, which were not completely resolved through fine-tuning.

This similarity could have resulted from the fundamental resemblance in their atomic and electronic structure.\cite{splendiani2010emerging, chhowalla2013chemistry}. However, similarities in the bonding and electronic properties are also observed between \ce{MoSe2} and \ce{WSe2} (Table S1), which are rarely misclassified in our analysis. In addition, the TMDs are grown at diverse conditions that are not necessarily optimum, and hence significant diversity in the features of the same material (Figure~\ref{fig:data}). Therefore, the morphological features and the classification performance result from the combination of the factors due to the atomic composition and the growth conditions in our training data. Interestingly, the models present such an excellent result despite this mix of factors.

\subsection{Expanding Models for New TMDs}
\begin{figure}
     \centering
     \includegraphics[width=.9\textwidth]{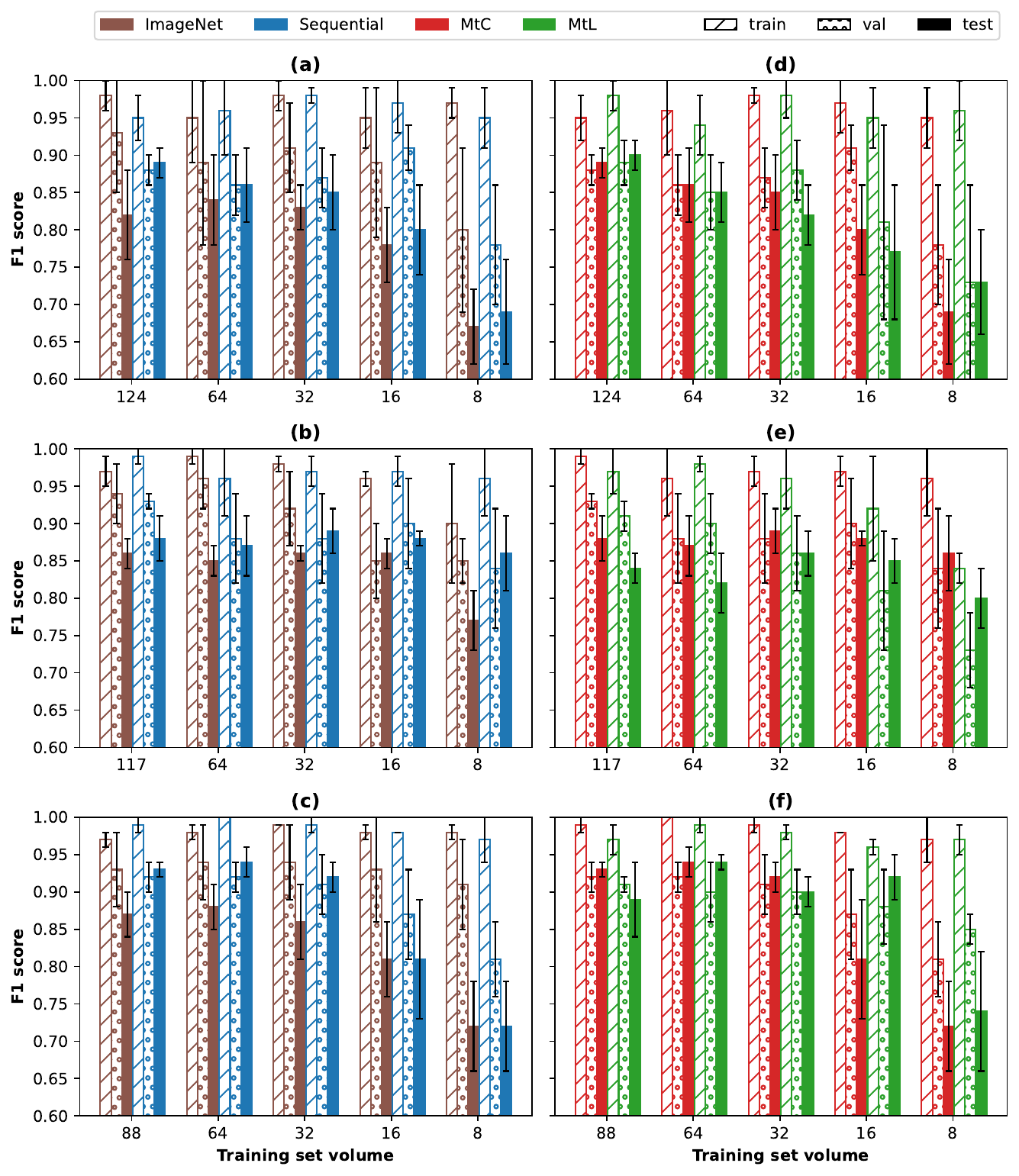}
         \caption{The F1 score of fine-tuning the ResNet model starting with the ImageNet weights (ImageNet) or continuing with the previous fine-tuned weights from fewer TMD classes (for instance, fine-tuning on three TMDs classification using the fine-tuned weights of two TMDs classification) (Sequential). The panels on the right (d-f) are for Sequential with multiclass (MtC) and multilabel (MtL) representations of the targets. Rows correspond to different numbers of TMDs being classified: a and d include three; b and e include four; c and f include five. The first training set volume is the total training set, while the subsequent volumes are the reduced sets. The results are the average predictions of the different data sets using the five models from the five-fold cross-validation.}
         \label{fig:f1score}
\end{figure}

As observed in our data, a balanced data set where all the classes of materials are equally represented is a rare occurrence in materials science (Table~\ref{tab:data}). The weighted random sampling we have adopted has significantly mitigated the effect of the imbalanced data. However, the imbalance in the data could be more pronounced than we have in our current data set. It is, therefore, important to explore how best to use transfer learning in such instances. Additionally, it is important to understand how transfer learning could be best adopted when models have to be trained to accommodate additional materials. For instance, if we have a model trained on \ce{MoS2} and \ce{WS2}, and additional TMD, say \ce{WSe2}, are grown. One option is to train different models from scratch or use the pre-trained weights of the model previously trained on the first two materials. Therefore, we explore the two options to determine which to adopt in such a scenario.

For three, four, and five TMDs classifications, we observe the effect of training our model by fine-tuning the ImageNet weight every time (ImageNet) or using sequential training where the weight from the previous model obtained from fewer materials classifications (Sequential) is fine-tuned (Figure~\ref{fig:schemes}). The use of Sequential is better in generalization to the held-out data or is similar to ImageNet in some instances (Figure~\ref{fig:f1score}). This shows that, with the additional class of TMDs, the weights from the fewer classes (Sequential) maximize the benefit of transfer learning in multi-material classifications. The difference could be very significant and holds for larger training volumes such as 124 in Figure~\ref{fig:f1score}a where test set F1 score for ImageNet and Sequential is 0.82 and 0.89, respectively, and smaller training volumes such as 8 in Figure~\ref{fig:f1score}b where test set F1 score for ImageNet and Sequential is 0.77 and 0.86, respectively.

The Sequential learning is examined for the MtC, and MtL target representations (Figure~\ref{fig:f1score}d-f). These give statistically indistinguishable results in most instances. However, for more material classes like the five in Figure~\ref{fig:f1score}f, MtL gives significantly better results with F1 scores of 0.92 versus 0.81 for MtC, for the 16 training volumes. This implies that MtL will likely perform better than MtC with increasing materials, particularly for smaller data volumes, which is useful in the context of new materials discovery.

\subsection{Model Explanation}
\begin{figure}
     \centering
     \includegraphics[width=\textwidth]{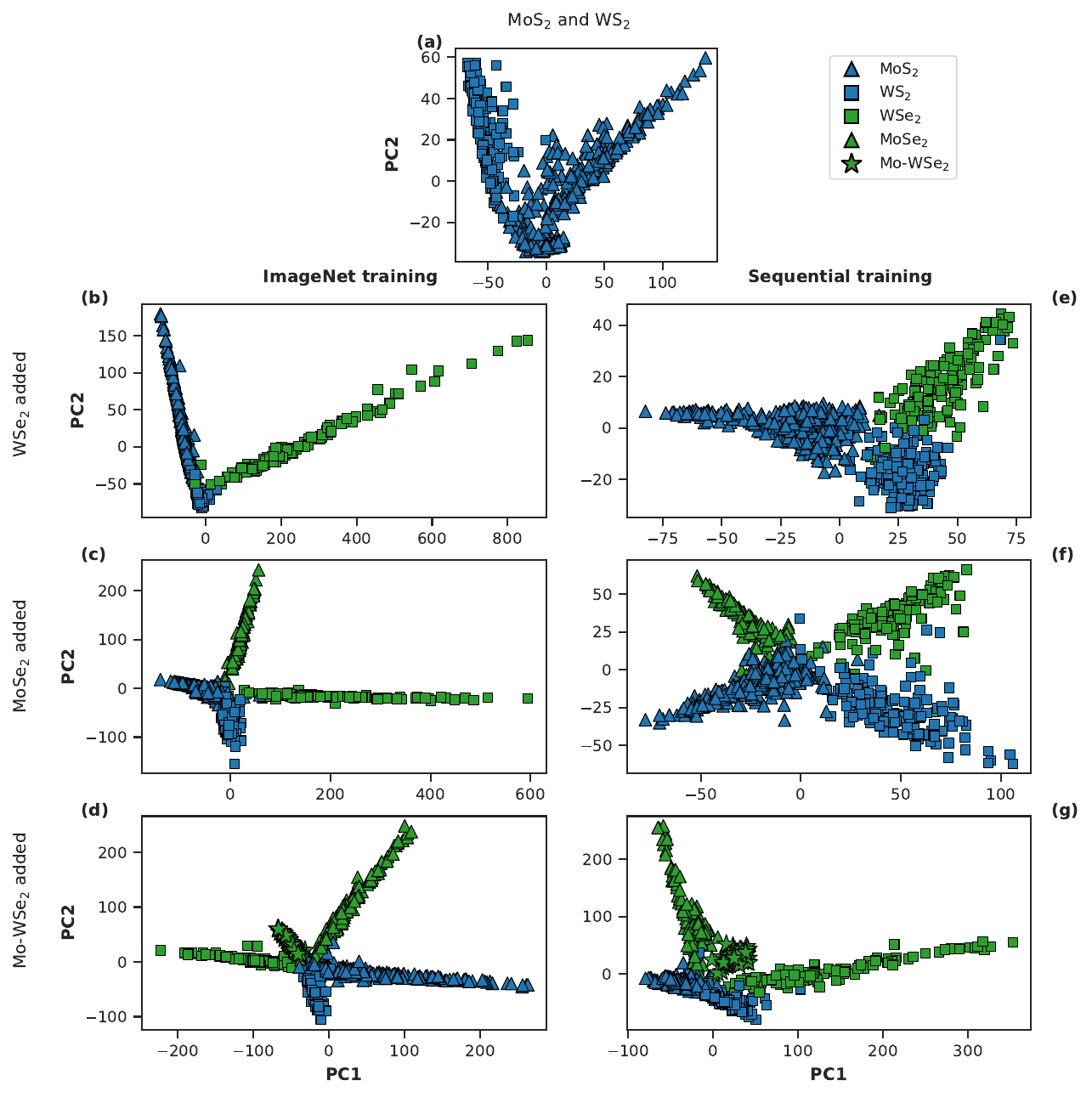}
         \caption{The scatter plot of the first two components of PCA of image features obtained from the trained model.
         Columns correspond to the training strategy: panels b-d are for the ImageNet training (models are based on the fine-tuning of the ImageNet weight), while e-g are for Sequential training (models are based on the sequential fine-tuning of the models trained on fewer classes. 
         For instance, three-material classification is based on fine-tuning the two-material classification model, and so on).
         Rows correspond to different numbers of TMDs being classified: a includes two, b and e include three, c and f include four, and d and g include five.
         }
         \label{fig:pca}
\end{figure}

\begin{figure}
     \centering
     \includegraphics[width=\textwidth]{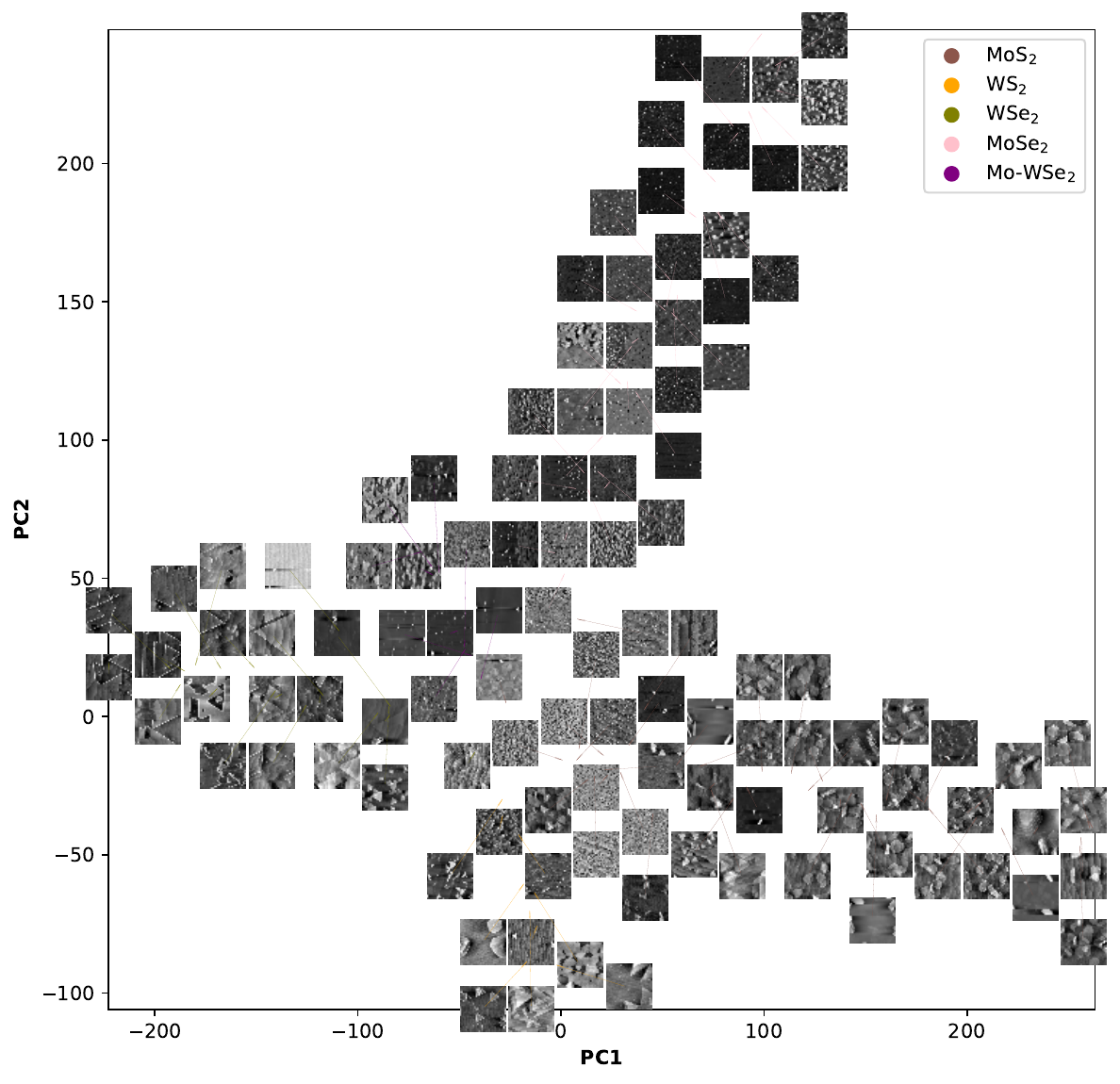}
         \caption{Sample images embedded in the PCA space shown in Figure~\ref{fig:pca}d. The arrow points to the actual image position in the embedding space.}
         \label{fig:pca_image}
\end{figure}

The high accuracy on the out-of-sample test set data underscores the significance of the discriminative latent features that the models associate with the different classes of TMDs. We therefore want to understand how the model learns in the different transfer learning modes, ImageNet and Sequential. Additionally, we want to gain insight into the visually observable image features that correspond to the latent discriminative features that the models associated with the different classes of TMDs. Using the second fully connected layer of the trained model, 100 features were extracted from the images. The models trained on two through five classes of TMDs were used to extract the image features. The principal component analysis, PCA, was applied to the features to obtain the first few components accounting for most feature variance. The data's latent space (first two components) is shown in Figure~\ref{fig:pca}.

In the Sequential training mode, it is interesting to observe that the different classes of the TMDs occupy relatively the same region even as more classes of materials are added. For instance, in the four and five TMDs classifications (Figures~\ref{fig:pca}f-g). The relative regions occupied by the \ce{MoS2}, \ce{WS2}, \ce{WSe2}, and \ce{MoSe2} in both the models classifying thefour TMDs and that for five, including \ce{Mo-WSe2}, are the same. The added fifth TMD in Figure~\ref{fig:pca}g only slid in between the \ce{MoS2} and \ce{WSe2}. The maintenance of the relative position by the different classes of the TMD, when more classes are added in Sequential, could indicate that the model retains the weights peculiar to classifying a set of TMDs even as it learns new weights for the added materials. This can lead to better generalization when more material classes are added.

It is also worth noting that the plots show distinct positions occupied by the different TMDs in the latent space. Interestingly, tremendous information in the AFM images could be captured in the 2D space so much that the materials classes could occupy a distinct position. Beyond the materials classes, it is observed that the models can associate a given atom with unique latent features irrespective of its present TMD. For instance, in Figure~\ref{fig:pca}d, a horizontal line could be drawn to separate the region containing TMDs with only \ce{S} as the constituent chalcogen from the rest, with the former occupying the lower part. Similarly, a slightly rotated diagonal would separate the region with TMDs containing only \ce{Mo} as the transition metal from others.

Beyond associating the different chemical compositions and TMDs to the unique regions of the 2D space, it is important to uncover potential physically meaningful features of the image that models use in distinguishing the classes of materials. To this end, the images are embedded in the latent space to show how the morphological features vary across the components and the materials (Figure~\ref{fig:pca_image}). Small domain sizes can be associated with the images located in the first quadrant of the space, mainly consisting of \ce{MoSe2}. In contrast, the images in the third and fourth quadrants mostly have bigger sizes of domains. However, it is difficult to visually determine how the image features vary within the classes in the space. Similarly, class activation maps (CAM) show the domain boundary and multilayer region being activated for the TMD classes (Supporting Information). However, the activations are not as discriminative as similar regions are activated irrespective of class (Figure S2 to S11).

\subsection{Latent and Physical Features}
\begin{table}[ht]
	\centering
	\caption{The image properties investigated for correlation with the PCA components obtained from the image features extracted using the trained model. The maximum value of negative and positive correlation coefficients between a given property and the component that gives the maximum value is shown. The rows in boldface are the first three properties with maximum correlation. The coefficient of correlation from the regression is also shown in the last column.} 
	\label{tab:property}
	\begin{tabular} {l |c |  c   c    |c c | c} \hline
	\multicolumn{2}{c|}{\multirow{2}{*}{\textbf{Property}}} & \multicolumn{2}{c|}{\textbf{Negative Correlation}} & \multicolumn{2}{c|}{\textbf{Positive Correlation}} & \textbf{Regression}\\ \cline{3-7}
   \multicolumn{2}{c|}{}	 &      R  & component    &   R & component & R\\ \hline
	1 & grain size  & -0.11        &  3 &  0.19         & 4    & 0.25  \\
 2  &  step edges  &  -0.37     &   7    &  0.15    & 3 & 0.47 \\
 3  &  anisotropy    &   -0.45  &  7    &  0.17 & 4 &  0.51  \\
         4  &  homogeneity  &  -0.48    &  7    &  0.17 &  4    &  0.55      \\
          5  &  energy       &  -0.31    &  7    &  0.15 &  10    & 0.38       \\
          6  &  correlation  &  -0.33    &  2    &  0.25 &  3    & 0.57       \\
          7  &  ASM          &  -0.14    &  7    &  0.10 &  10    & 0.19       \\ 
	\bf{8} & \bf{grain density} & \bf{-0.25}     &   \bf{5}    &   \bf{0.48}     & \bf{7}    & 0.61 \\
    9 &  \bf{DoG blob}     &  -0.29    &  4    &  0.42 &  7 &     \bf{0.65}       \\ 
	10  &  rms roughness&  -0.27     &   4   &  0.43    & 7 & 0.61  \\
	11  &  defect density    &  -0.20     &  3    &  0.21 & 7   &  0.39  \\
	\bf{12}   &  \bf{local variation}  &  \bf{-0.25} &   \bf{4}     &   \bf{0.50} & \bf{7}  &  \bf{0.65}\\ 
	13   &   contrast &  -0.25  &    3    &    0.35 & 7     & 0.59   \\
        \bf{14}   &  \bf{dissimilarity}    & \bf{-0.28} &  \bf{3}    &  \bf{0.48} &  7   & \bf{0.66}        \\\hline
       
		\end{tabular}
	\end{table}
Following the challenge of visually discerning the discriminative image properties that the models use in classifying the TMDs, a detailed search of a relationship between the latent features and quantified physically meaningful image features (physical features) was carried out. The material images were analyzed using several metrics, including microstructural features such as the grain size, grain density,  and Difference of Gaussian (DoG) blob, topographical and defect-related features such as the surface roughness, step edges, defect density, anisotropy, and local variation and the Gray-Level Co-occurrence Matrix (GLCM) features such as the contrast, dissimilarity, homogeneity, energy, correlation, and angular second moment (ASM) (Table \ref{tab:property}). The magnitude of each physical feature was normalized by dividing them by the maximum values such that they all fall into the 0 to 1 range.

\begin{figure}
     \centering
     \includegraphics[width=\textwidth]{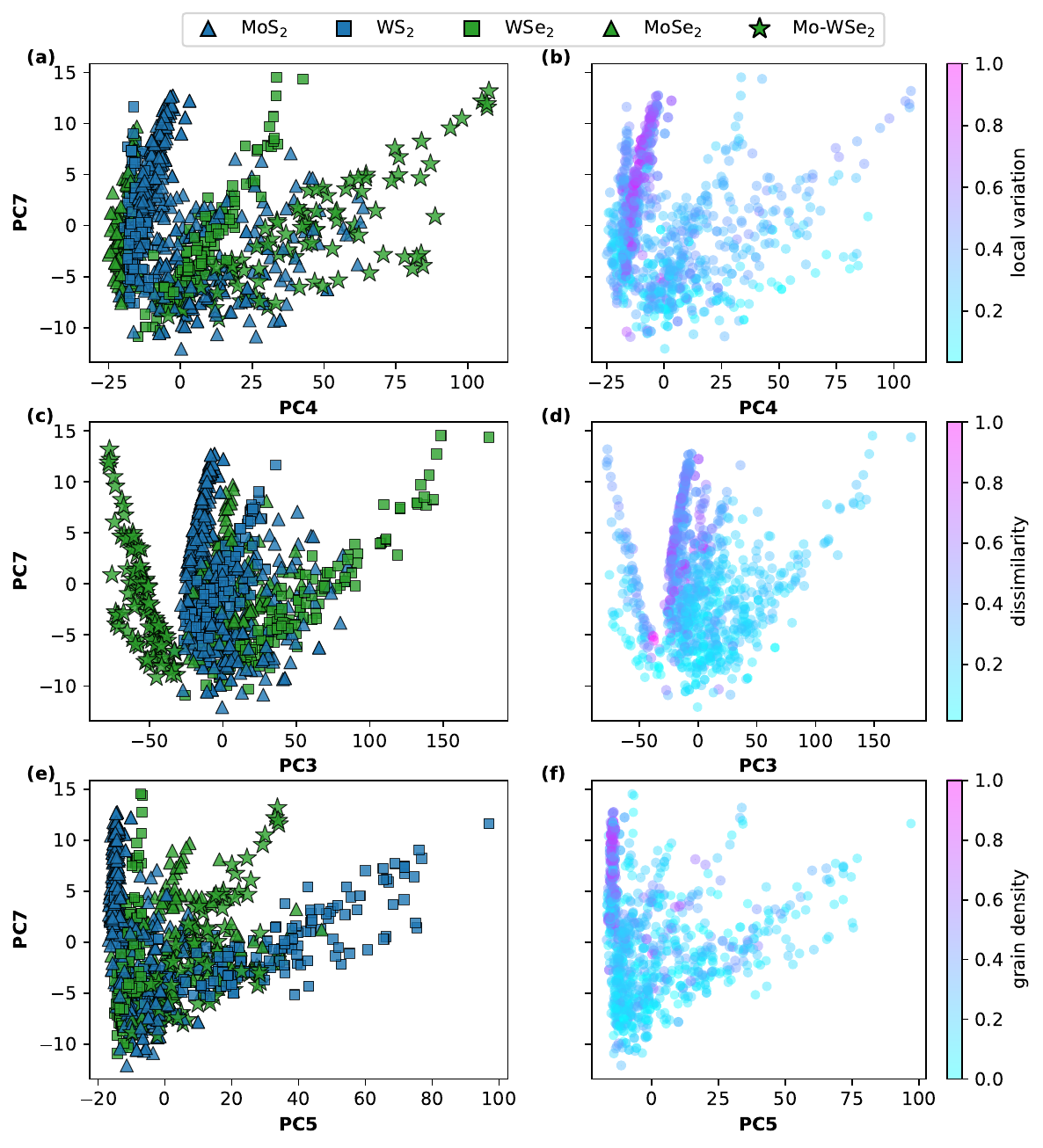}
         \caption{The scatter plot of the data point (left) and the image properties (right) in two PCA components (of the features extracted from the trained model) with the highest correlation with image properties. Component 6 (PC6) has the highest positive correlation with the properties while PC1, PC4, and PC4 have the highest negative correlation with local variation, rms roughness, and DoG blob, respectively. The data is on the left (a, c, and e), and the corresponding image property is on the right (b, d, and f).}
         \label{fig:property}
\end{figure} 
After quantifying the physical features, the Pearson correlation coefficient between each quantity and the dimensions of the image's latent features (PCA components of the features extracted using the trained model) was obtained. We determined which dimension of the latent feature has the greatest magnitude of the correlation coefficient with the physical features. The analysis shows the physical features with the highest correlation with any dimension of the latent features are grain density, local variation, and dissimilarity with positive correlation coefficients of 0.48, 0.50, and 0.48, respectively. The seventh dimension of the latent features correlates with the three physical features, while minimal negative correlation is observed between the fifth dimension and grain density, the fourth and local variation, and the third and dissimilarity (Table \ref{tab:property} and Figure \ref{fig:property}).
\begin{figure}
     \centering
     \includegraphics[width=\textwidth]{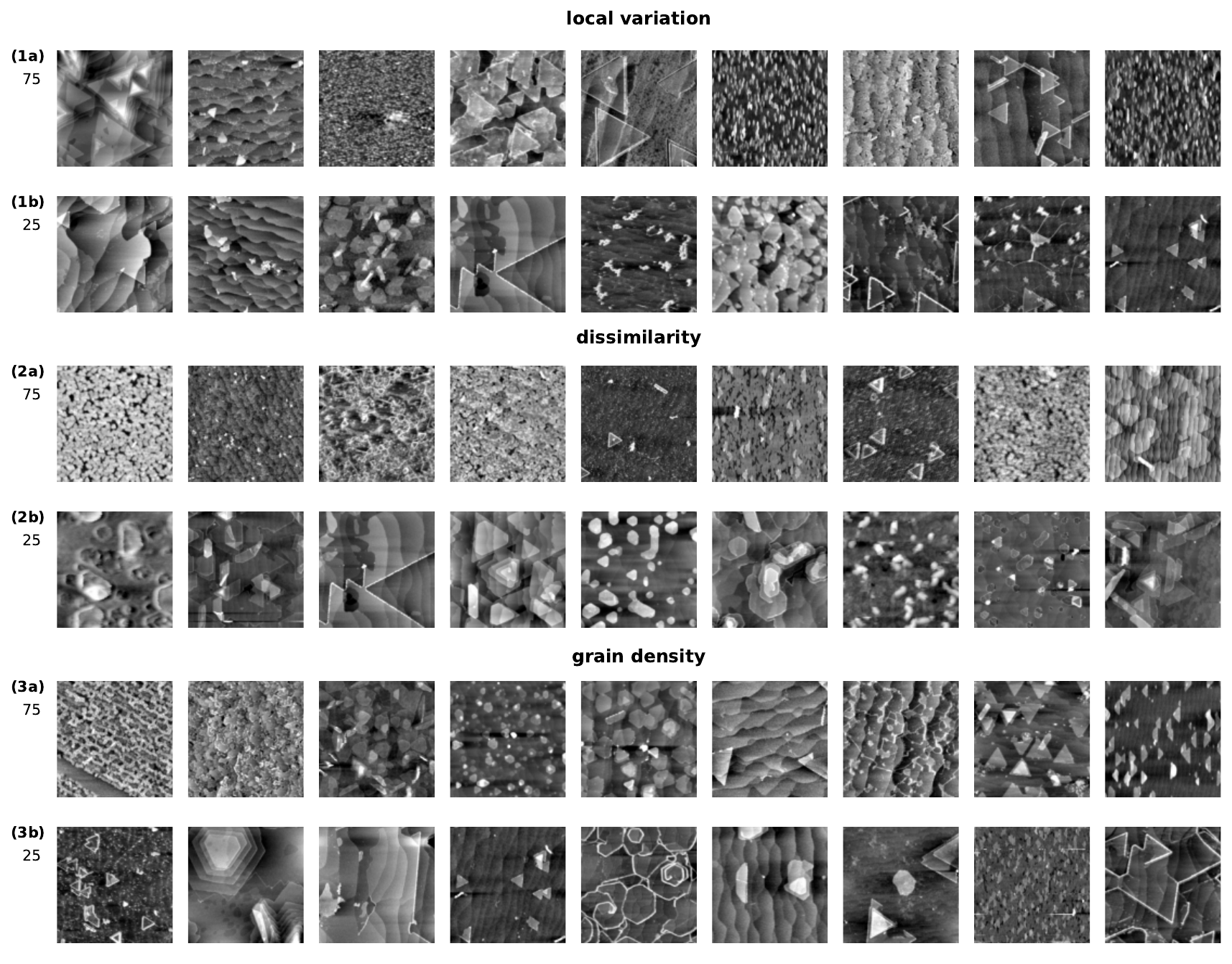}
         \caption{The sample images with the high (at least in the 75th percentile) and low (in the 25th percentile and less) values of image properties; degree of local structural variation (local variation), dissimilarity, and grain density. The first rows (1a, 2a, and 3a) show images with the highest value of the quantities while the following rows (1b, 2b, and 3b) show the images with the lowest value of the quantities.}
         \label{fig:prop_img}
\end{figure}
For the quantities with the highest correlation, sample images with the high magnitude (75th percentile and upward) and those with the low values (25th percentile and less) are shown in Figure \ref{fig:prop_img}.

For more insight into the relationship between the physical and latent space, we have obtained the cosine similarity between each of the physical features and the components of the latent features. This is to especially unveil physical features that are orthogonal to the latent features. This result is presented in a heat map in supporting information (Figure S13). As expected, the highest values are observed for grain density, contrast, dissimilarity, and contrast with the PC7, in agreement with the Pearson correlation. Most of the physical features are (nearly) orthogonal with some of the dimensions of the latent features (cosine similarity $\approx$ 0). PC1, for instance, has only minimal deviation from orthogonality with all of the physical features. 

To determine how these properties vary across the classes of TMDs in the latent space, a scatter plot of the data points in dimension with the negative correlation and that with the positive correlations were obtained (Figure \ref{fig:property}). Additionally, the quantities of the physical features were embedded in the latent space. As expected, unlike the first two components (shown in Figure~\ref{fig:pca}), these components do not show a distinctive clustering of the different classes of the TMDs because of their relatively low explained variance. However, we could notice a group of \ce{MoS2} and \ce{WS2} have the highest values of the local variation and dissimilarity, and \ce{MoS2} with the highest grain density. 

To further explain how the latent and physical features are related, we went beyond the correlation to do a regression analysis between the features. As expected, this results in a correlation coefficient of about 0.7, up from the maximum value of 0.5 obtained from the Pearson correlation (Figure~\ref{tab:property}). Features with the highest correlation with the latent features include the DoG blob, local variation, and dissimilarity. 

\begin{figure}
     \centering
     \includegraphics[width=\textwidth]{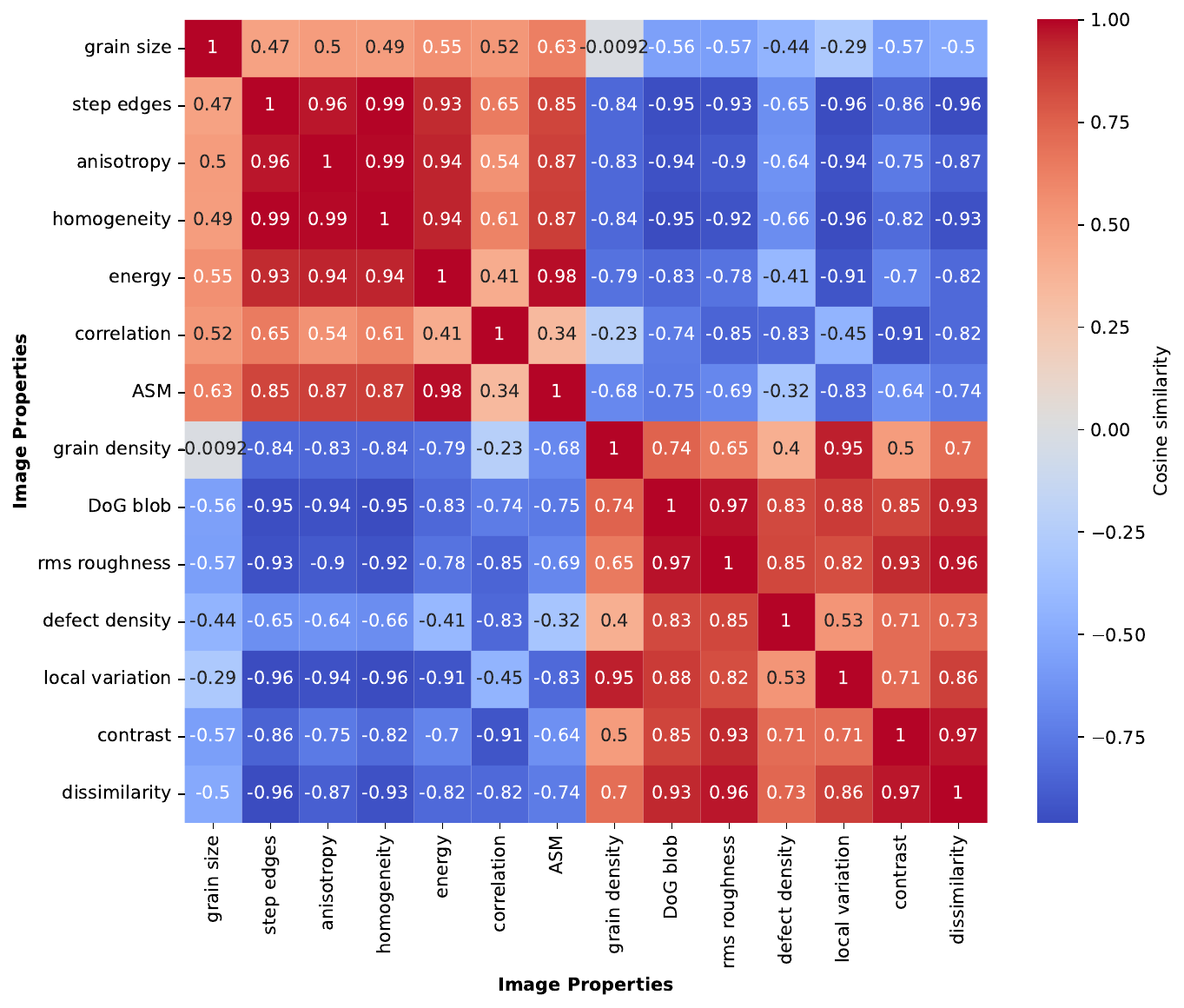}
         \caption{The Cosine similarity among the regression directions of the physical properties. Regression was obtained between the physical properties calculated from the AFM images and the ten first principal components (from PCA) of the image features extracted using the trained model. Similarity values of 1, -1, and 0 indicate parallel, antiparallel, and orthogonal directions, respectively.}
         \label{fig:similarity}
\end{figure}
The relationship among the regression direction of physical features is analyzed using the Cosine similarity (Figure~\ref{fig:similarity}). This analysis shows diverse directions and some groupings among the features. For instance, using local variation as the standard, properties such as grain density, DoG blob, and dissimilarity have a Cosine similarity of about 0.9, revealing almost identical regression direction. In contrast, properties such as step edges, anisotropy, and homogeneity show high dissimilarity (negative values indicating opposite direction) with the local variation. The embedding of our data in the representative regression directions similar and dissimilar to the local variation (dissimilarity and homogeneity) show features, for instance, local variation, dissimilarity, and grain density, with high and low values on the right upper, and left lower region, respectively (Figure S12). 
Interestingly, despite most of these properties exhibiting high correlation values with the latent features, their regression directions are diverse. This indicates that the model's discriminative latent features could be related to a fusion of several physical features.

\section{Conclusion}
We present deep learning models trained on AFM images of five classes of metal-organic chemical deposition (MOCVD) grown TMDs; \ce{MoS2}, \ce{WS2}, \ce{WSe2}, \ce{MoSe2}, and \ce{Mo-WSe2}.
Classification based on two labeling schemes, multiclass (MtC) and multilabel (MtL), were investigated in the context of several transfer learning strategies to accommodate imbalanced and low-data scenarios.
We also analyzed the discriminative latent features used by the ML models to classify materials and their correlation with physical features. 

Our models present accurate out-of-sample predictions. The MtC and MtL both perform excellently, with the MtL showing minimal improvement over the MtC labeling. While the MtC classification gives validation and test accuracies of about 94 and 87\%, respectively, the MtL gives 95 and 89\%, respectively. This is an encouraging result since each labeling scheme has advantages in different contexts, and our results suggest that either can be used effectively. The models' explanation analyses suggest that their latent features used to distinguish one material from others are interrelated to a fusion of several physical features. The analysis shows the physical features with the highest correlation with any dimension of the latent features are grain density, local variation, and dissimilarity, with positive correlation coefficients of 0.48, 0.50, and 0.48, respectively, and complementary negative correlations. The transfer learning optimization modality and the exploration of the correlation between the models' latent features and the image's physical features provide important frameworks that can be applied to other classes of materials beyond TMDs to enhance the models' performance and explainability which can accelerate the inverse design of materials for technological applications.

\section*{Acknowledgments}
This study is based on research conducted at the Pennsylvania State University Two-Dimensional Crystal Consortium—Materials Innovation Platform (2DCC-MIP), supported by NSF cooperative agreement DMR-2039351.

\section*{Data Availability}
The raw data required to reproduce these findings can be downloaded from Ref\citenum{Moses_Reinhart}.
The processed data and code required to reproduce these findings can be downloaded from Ref\citenum{moses_2024_13961220}.

\bibliography{main}
\end{document}